\documentclass[12pt, conference, compsocconf]{IEEEtran}
\usepackage{graphicx}
\DeclareGraphicsExtensions{.eps}
\usepackage{subfig}
\usepackage{url}

\ifCLASSINFOpdf
\else
\fi
\begin{document}
%
\title{High Volume Computing: Identifying and Characterizing Throughput Oriented Workloads in Data Centers}


\author{\IEEEauthorblockN{Jianfeng Zhan, Lixin Zhang, Ninghui Sun, Lei Wang, Zhen Jia, and Chunjie Luo}
\IEEEauthorblockA{ State Key Laboratory of Computer Architecture\\
Institute of Computing Technology\\
Chinese Academy of Sciences\\
Beijing, China\\
Email: \{zhanjianfeng, zhanglixin, snh\}@ict.ac.cn, \{wl, jiazhen, luochunjie\}@ncic.ac.cn}
}


%


\maketitle

\begin{abstract}
For the first time, this paper systematically identifies three categories of throughput oriented workloads in data centers: services, data processing applications, and interactive real-time applications,  whose targets are to increase the volume of throughput in terms of \emph{processed requests or data}, or \emph{supported maximum number of simultaneous subscribers}, respectively, and we coin a new term  \emph{\underline{h}igh \underline{v}olume  \underline{c}omputing (in short HVC)} to describe those workloads and data center computer systems designed for them. We characterize  and compare HVC with other computing paradigms, e.g., high throughput computing,  warehouse-scale computing, and  cloud computing, in terms of levels, workloads, metrics, coupling degree, data scales, and number of jobs or service instances. We also preliminarily report our ongoing work on the metrics and  benchmarks for HVC systems, which is the foundation of designing innovative data center computer systems for HVC workloads.
\end{abstract}


%
\IEEEpeerreviewmaketitle

\section{Introduction}

\IEEEPARstart{I}n the past decade, there are three trends in computing domains. First, more and more \emph{services}, involving a large amount of data,  are deployed in data centers to serve the masses, e.g., Google search engine and Google Map.  Second, massive data are produced, stored, and analyzed in real time or off line. According to the annual survey of the global digital output by IDC, the total amount of global data passes 1.2 zettabytes  in 2010. In this paper, we call applications that produce, store, and analyze massive data \emph{data processing applications}, which is also referred to as big data applications.
Third, lots of  users tend to use streaming media or VoIP for fun or communications. Different from an ordinary Web server, a  VoIP application will maintain a user session of a long period (e.g., more than five minutes) while guaranteeing the \emph{real time} quality of service, which we call \emph{an  interactive real-time application}.

The workloads mentioned above  consist of a large amount of loosely coupled workloads instead of a big tightly coupled job. The nature of this class of workloads is throughput-oriented, and the target of data center computer systems designed for them is to increase the volume of throughput in terms of processed requests (for services), or
processed data (for data processing applications), or the maximum number of simultaneous
subscribers (for interactive real-time applications), performed or supported in
data centers.  So as to pay attention to this class of workloads and computer systems designed for them,  in this paper, we coin a new term \emph{\underline{h}igh \underline{v}olume \underline{c}omputing} (nature: throughput computing; target: high volume, in short HVC) to describe this class of workloads or data center computer systems designed for them.

In this paper, we identify three categories of workloads in HVC: \emph{services}, \emph{data processing applications}, and
 \emph{interactive real-time applications}, all of which are throughput-oriented workloads.   A service is a group of applications that collaborate
to receive user requests and return responses to end users. More and more emerging services are data-intensive, e.g., Google search engine or Google Map.
Data processing applications  produce, store, and  analyze massive data, and we only focus on \emph{loosely coupled} data processing applications, excluding tightly coupled data-intensive computing, e.g., those written in MPI.
 Typical examples are MapReduce or Dryad based computing. We also include \emph{data stream} applications that process continuous unbounded streams of data in real time into the second category
 of HVC workloads.
Different from an ordinary Web server,  an interactive real-time application will maintain a user session of a long period while guaranteeing the \emph{real time} quality of service. Typical interactive real-time applications include streaming media, desktop  clouds \cite{Desktop_cloud}, and  Voice over IP (VoIP) applications. The details of three categories of workloads can be found at \ref{HVC_defintion}.

Despite  several computing paradigms are not formally or clearly defined, e.g., \underline{w}arehouse-\underline{s}cale \underline{c}omputing (WSC) \cite{53-datacenter}, \underline{d}ata-\underline{i}ntensive \underline{s}calable \underline{c}omputing (DISC) \cite{DISC}, and cloud computing, we compare HVC with several computing paradigms in terms of six dimensions:
levels, workloads, metrics, coupling degree, data scales, and number of jobs or service instances as shown in Table \ref{Characterizing_computing_paradigms}.
Our definition of HVC is towards data center computer systems, while many task computing \cite{1raicu2008many} and high throughput computing \cite{HTC} are defined towards runtime systems.
In terms of workloads and respective metrics, both high throughput computing and high performance computing are about scientific computing centering around floating point operations,
while most of HVC applications have few floating point operations as uncovered in our preliminary work \cite{Search_Benchmark}.
Meanwhile, we also notice that many emerging workloads can be included into one or two categories of HVC workloads,
  e.g., WSC (into the first and second categories), and DISC (into the second category). As for (public) cloud computing, we believe it is basically a business model of renting computing or storage resources,
  which heavily relies upon virtualization technologies, however HVC is defined in terms of workloads.
  We think many well-know workloads in cloud \cite{ClearingCloud} can  be included into HVC, but HPC in cloud workloads \cite{DawningCloud}  \cite{HPC_in_Cloud} are excluded, since they are   tightly-coupled. 
%

After widely investigating previous benchmarks, we found there is no systematic work on benchmarking data center computer systems in the context of our identified three categories of throughput-oriented workloads. 
We present our preliminary work on the  metrics and benchmarks for HVC systems.


The remainder of the paper is organized as follows. Section \ref{different_paradigms} characterize and compare different computing paradigms.  Section \ref{benchmark_HVC} revisits previous benchmarks and report our preliminary work on the HVC metrics and benchmarks.  Section \ref{conclusion} draws a conclusion.

\section{Characterizing computing paradigms} \label{different_paradigms}

\begin{table*}[hbtp]
\renewcommand{\arraystretch}{1.3}
\caption{Characterizing different computing paradigms.}
\label{Characterizing_computing_paradigms}
\centering
\begin{tabular}{|p{3.0cm}|p{3.0cm}|p{2.5cm}|p{2.5cm}|p{1.0cm}|p{1.0cm}|p{1.8cm}|}
  \hline
  \itshape Computing paradigm &\itshape level &\itshape Workloads & \itshape Metrics & \itshape  Coupling degree  & \itshape Data scale &\itshape $\sharp$ jobs or service instances\\ \hline
  \itshape High performance computing &Super computers & Scientific computing: heroic MPI  applications &Float point operations per second & Tight &n/a  &Low\\ \hline
      \itshape High performance throughput computing \cite{High_performance_throughput_computing} & Processors& Traditional server workloads & Overall work performed
over a fixed time period &loose &n/a & Low  \\ \hline
  \itshape High throughput computing \cite{HTC} & Distributed runtime systems & Scientific computing& Float point operations per month &loose &n/a &Medium\\ \hline
  \itshape Many task computing \cite{1raicu2008many} & Runtime systems & Scientific computing or data analysis: workflow jobs & Tasks per second &Tight or loose &n/a &Large \\ \hline
    \itshape Data-intensive scalable computing \cite{DISC} or data center computing \cite{HotOS} & Runtime systems & Data analysis: MapReduce-like jobs & n/a &Loose &Large &Large\\ \hline
      \itshape Warehouse-scale computing \cite{53-datacenter} &Data centers for Internet services, belonging to a single organization &Very large Internet services &n/a &Loose &large &Large \\ \hline
        \itshape Cloud computing \cite{Cloud_defintion} \cite{2armbrust2009above} & Hosted data centers & SaaS + utility computing & n/a &Loose &n/a &Large \\ \hline
         &  & Services &Requests per minutes and joule &Loose  &Medium &Large \\
            \itshape \textbf{\underline{H}igh \underline{v}olume  \underline{c}omputing (HVC)} & Data centers & Data processing applications &Data processed per minute and joule   &Loose &Large &Large  \\
             &  & Interactive real-time applications &Maximum number of simultaneous subscribers  and subscribers per watt  &Loose &From medium to large &Large \\ \hline
\end{tabular}
\end{table*}

In this section, we give out the definition of HVC, and identify its distinguished differences from other computing paradigms.

\subsection{What is HVC?} \label{HVC_defintion}

HVC is a data center based computing paradigm focusing on throughput-oriented workloads. 
The target of a data center computer system designed for HVC workloads is to increase the volume of throughput in terms of \emph{requests}, or \emph{processed data}, or {\em the maximum number of simultaneous subscribers}, which are performed or  supported in a data center.

 In Table \ref{Characterizing_computing_paradigms}, we characterize HVC from six dimensions: levels, workloads, metrics, coupling degree, data scales, and number of jobs or service instances.

 The HVC system is defined on a data center level. We identify three categories of workloads in HVC: \emph{services}, \emph{data processing applications}, and
 \emph{interactive real-time applications}.
 Services belong to the first category of HVC workloads. A service is a group of applications  that collaborate to receive user requests and return responses to  end users.
 We call a group of applications that independently process requests \emph{a service instance}.  For a large Internet service, a large amount of service instances are deployed with requests distribution enabled by load balancers.
 Since each request is independent, a service in itself is loosely coupled.  For an ordinary Apache Web server, the data scale is lower, while for a search engine provided by Google, the data scale is large.
 More and more emerging services are data-intensive.

 The second category of HVC workloads is \emph{data processing applications}. Please note that we only include loosely coupled data-intensive computing, e.g., MapReduce jobs,  into HVC, \emph{excluding data-intensive MPI applications}. 
 In running, MapReduce tasks are independent, significantly different from batch jobs of programming models like MPI in which tasks execute concurrently and communicate during their execution \cite{Quincy}. The data scale of this category of workloads is large, which are also referred to as \emph{big data applications},  and hence will produces large amount of tasks.  We also include data stream applications into this category of HVC workloads.  For example, S4 is a platform that allows programmers to easily develop applications for processing continuous unbounded streams of data \cite{S4}.

 The third category of HVC applications is \emph{interactive real-time applications}. Different from an ordinary Web server, an interactive real-time application will maintain a user session of a long period while guaranteeing the \emph{real time} quality of service. Typical interactive real-time applications include streaming media--- multimedia that is constantly delivered to an end-user by a service provider, desktop  clouds \cite{Desktop_cloud}, and  Voice over IP (VoIP) applications.
 For this category of applications, the workload is loosely coupled because of independent requests or desktop applications; the data scale varies from medium to large, and the number of tasks or service instances is large. 

\subsection{Identifying differences of HVC}

\subsubsection{High performance computing} There are two-fold differences of HVC from \underline{h}igh \underline{p}erformance \underline{c}omputing (in short, HPC): first, workloads are different. HPC is mainly about scientific computing and usually a large-scale  heroic MPI application, which is tightly coupled, while HVC is loosely coupled and commonly composed of a large amount of jobs or service instances. Second, the metric is  different. The metric of HPC is \emph{floating point operation per second}. However, in HVC, most of workloads, e.g., web search engines reported in \cite{Search_Benchmark},  have few float point operations.

\subsubsection{High throughput computing} Livny {\em et al.} refer to the environments that can be deliver large amounts of processing capacity over very long periods of time as high throughput computing \cite{HTC}.
There are three-fold differences of of HVC:
first,  high throughput computing is defined on the level of distributed runtime systems, while HVC is defined on the level of data center computer systems;
 second, the workloads of high throughput computing are towards scientific computing, while our HVC includes three categories of applications; third, the metric of high throughput computing
  is floating point operations per month or year. However, in HVC, major workloads, e.g., web search engines \cite{Search_Benchmark}, have few float point operations.

\subsubsection{Many task computing} According to \cite{1raicu2008many}, many task computing differs from high throughput computing in the emphasis
of using large number of computing resources over short periods of time to accomplish many
computational tasks 
where primary metrics are measured in seconds (e.g. FLOPS, tasks/sec, MB/s I/O
rates), as opposed to operations  per month.
In terms of workloads, many task computing denotes high-performance
computations comprising multiple distinct activities,
coupled via file system operations \cite{1raicu2008many}. 
The set of tasks may be
loosely coupled or tightly coupled.

The differences of HVC are as follows: first, the concept of many task computing is towards runtime systems, while our HVC is on the level of data center computer systems; second, the workloads and respective metrics are different.
Many task computing includes tightly or loosely coupled applications from scientific computing or data analysis domains, while HVC includes three categories of  loosely coupled applications. 

\subsubsection{High performance throughput computing} Chaudhry {\em et al. } \cite{High_performance_throughput_computing} call the combination of
\emph{high-end single-thread performance} and \emph{a high degree of multicore, multithreading}
 high performance throughput computing. According to \cite{High_performance_throughput_computing},
systems designed for throughput computing emphasize the the aggregate amount of computation
performed by all functional units, threads,
 cores,  chips,  coprocessors  and
 network interface cards in a system over a fixed time period as opposed to focusing
on a speed metric describing how fast a single
core or a thread executes a benchmark.

Different from HVC, high performance throughput computing is defined on the level of processors, targeting traditional server workloads, e.g., database workloads, TPC-C, SPECint2000, SPECfp2000 exemplified in \cite{High_performance_throughput_computing}, 
while HVC is defined on the data center level.  Different from high performance throughput computing, we have identifies three categories of workloads in HVC.
In terms of workloads, high performance throughput computing mainly focuses on the first category of HVC applications (excluding data-intensive services like search engines) in addition to float point operation-intensive benchmarks like SPECfp2000. Moreover, we define different metrics for three different categories of HVC applications in \ref{our_work}.

\subsubsection{\underline{D}ata-\underline{i}ntensive \underline{s}calable \underline{c}omputing (in short, DISC) or data center computing}  Bryant {\em et al.} did not formally define what is DISC \cite{DISC}. Instead, they characterized DISC through comparing it with cloud computing and high performance computing in \cite{DISC}: first, cloud computing is towards hosted services, e.g., web-based Email services, while DISC refers to very large, shared data repository enabling complex analysis \cite{DISC}; second, from a perspective of programming model, a HPC program is described at very low level with specifying detailed control of
processing  communications, while DISC applications are written in terms of high-level operations on data, and the runtime system controls
scheduling and  load balancing \cite{DISC}. In general, a DISC application is written in the MapReduce programming model, which splits jobs into small tasks that are run on the cluster¡¯s compute nodes.
No one formally defines what is data center computing. In \cite{HotOS}, it refers to computing performed with frameworks such as MapReduce, Hadoop, and Dryad. Through these frameworks, computation can be performed
on large datasets in a fault-tolerant way, while hiding the complexities of the distributed nature of the
cluster \cite{HotOS}.  According to \cite{HotOS}, data center computing shares the same view like that of DISC.

Basically, DISC or data center computing can be included into the second category of HVC workloads. In addition,  our HVC is defined toward data center computer systems instead of MapReduce-like runtime systems mentioned in \cite{DISC} \cite{HotOS}.
Second, Bryant {\em et al.} do not formally define the metrics, while we define three different metrics in evaluating data center computer systems for three categories of applications in \ref{our_work}.

\subsubsection{Warehouse-scale computing} According to \cite{53-datacenter}, the trend toward server-side computing and the exploding popularity of Internet services has created a new class of computing systems that  Hoelzle {\em et al.} have named warehouse-scale computers, or
WSC. WSC is meant to call attention to the most distinguishing feature of these machines:
the massive scale of their software infrastructure, data repositories, and hardware platform \cite{53-datacenter}. 
WSC demonstrates the following characteristics \cite{53-datacenter}:
belonging to a single organization, using  a relatively homogeneous hardware and system software platform,
and sharing a common systems management layer \cite{53-datacenter}. Most importantly, WSCs run a smaller number of very large Internet services \cite{53-datacenter}.

There are two differences of HVC from WSC: first, WSC is towards data centers for Internet
services, belonging to a single organization, while in HVC, many small-to-medium scale services from different organization may be hosted in the same data center, the aggregate number of which is large. Second, in terms of workloads, WSC can be included in HVC,
meanwhile HVC covers more categories of workloads, e.g.,  interactive real-time applications. Moreover, we have given out the specific metrics for different categories of workloads,
while in WSC, no metric is defined in \cite{53-datacenter}.

\subsubsection{Cloud computing}  Vaquero {\em et al.} \cite{Cloud_defintion} defined cloud as a large pool
of easily usable and accessible virtualized resources (such as hardware, development platforms and/or
services). 
According to the widely cited Berkeley's technology report \cite{2armbrust2009above}, 
when a cloud is made available in a pay-as-you-go manner to the general public,  it  is called a public cloud; the service being sold is
utility computing \cite{2armbrust2009above}. The term private cloud is used  to refer to internal data centers of a business or other organization,
not made available to the general public \cite{2armbrust2009above}, which is also characterized as warehouse-scale computing  in \cite{53-datacenter} or data center computing in  \cite{HotOS}. 

Basically, we believe that (public) cloud computing is a fine grain pay-per-use business model of renting computing or storage resources, which heavily relies upon virtualization technologies, while our HVC is defined in terms of workloads. Though virtualization technologies indeed bring changes to workloads, which is worth further investigation, however, well-known workloads in cloud studied in \cite{ClearingCloud}, e.g., \emph{NoSQL} data serving, MapReduce, Web search,  can be included into HVC.   A new trend is that practitioners in HPC  communities advocate an innovative computing paradigm---HPC in cloud \cite{DawningCloud} \cite{HPC_in_Cloud}, which suggests renting virtual machines from a public cloud for running parallel applications. Since HPC in cloud workloads are tightly coupled, we exclude them from HVC.  Moreover, in the context of a business model, no one formally defines the metrics for evaluating cloud from perspectives of both hardware and software systems.

\subsection{Discussion} Despite  several computing paradigms are not formally or clearly defined, e.g., DISC, WSC, and cloud computing,
we characterize several new computing paradigms 
 and compare the definition of our HVC from them.

We draw three conclusions in this Section: first, the definition of HVC is towards data center computer systems, while many task computing, high throughput computing, or DISC  are defined towards runtime systems. Second, in terms of workloads and respective metrics, both high throughput computing and high performance computing are about scientific computing centering around floating point operations, while most of HVC applications have few floating point operations.  Third, many emerging workloads can be included into one or two categories of HVC workloads  e.g., WSC (into the first and second categories), DISC or data center computing (into the second category); With the exception of HPC in cloud workloads\cite{DawningCloud},  well-know workloads in cloud studied in \cite{ClearingCloud} can be included into HVC. Moreover, (public) cloud computing is basically a business model of renting computing or storage resources, while our HVC is defined in terms of workloads.



\section{Benchmarking  HVC systems} \label{benchmark_HVC}

In this section, we revisit previous benchmarks, and present our current work on HVC metrics and benchmarks.

\subsection{Revisiting previous benchmarks and metrics}



%

Table \ref{Different_Benchmarks_metrics} summarizes different benchmarks and their respective levels, workloads,  and metrics.

The LINPACK benchmark report \cite{LINPACK}
describes the performance for solving a general dense matrix problem Ax = b. 
 Performance is often measured in terms of floating point operations per
second (flop/s). 
Focusing on evaluating the ability of traversing the whole memory of the machine \cite{betterBechmark}, Murphy {\em et al.} have
put together a benchmark they call Graph 500, and the metric is the traversed edges per second (TEPS).
Since energy efficiency becomes more and more important, the Green500 List ranks supercomputers used primarily for scientific production codes according to the amount of power needed to complete a fixed amount of work \cite{Green500}. 
Sun Microsystems also proposed the SWaP (space, watts, and
performance) metric \cite{SUN_SPACE_WATTS} to evaluate enterprise systems from perspectives of both data center space efficiency and power consumption.


Rivoire {\em et al.} \cite{Green_Model_Metrics} proposed the JouleSort benchmark. The external sort from the sort benchmark specification
\url{(http://research.microsoft.com/research/barc/SortBenchmark/default.htm)}
is chosen for the benchmark¡¯s workload.  The metrics is records sorted per Joule. SPECpower\_ssj2008 is the first industry-standard SPEC benchmark that evaluates the power and performance characteristics of volume server class  and multi-node class computers \cite{SPEC}. The initial benchmark addresses only  the performance of server side Java---SPECjbb2005.  The metric is  performance per watt metric in terms of ssj\_ops/watt.

TPC defined transaction processing and database benchmarks, the goal of which is to define a set of functional requirements that can be run on any transaction processing system, regardless of hardware or operating system \cite{TPC}.
Most of the TPC benchmarks are obsolete, and only three are still used: TPC-C, TPC-E, and TPC-H. TPC-C  centers around the principal activities (transactions) of an order-entry environment \cite{TPC}.
TPC-E models a brokerage firm with customers who generate transactions related to trades, account inquiries, and market research \cite{TPC}. Different from TPC-C and TPC-E, 
TPC-H   models the analysis end of the business environment where trends are computed and refined data are produced to support the making of sound business decisions \cite{TPC}.  For the TPC benchmarks, the metrics are application-specific. For example, the metric of TPC-C is the number of New-Order transactions executed per minute.

In the context of data center  computing, HiBench \cite{HiBench}, GridMix2 or GridMix 3 \cite{GridMix}, WL Suite \cite{WL Suite} is proposed to evaluate MapReduce runtime systems, respectively. The workloads are data analysis applications. The metrics are throughput in terms of the number of tasks per minute \cite{HiBench}, and job running time, widely used in
batch queuing systems \cite{10gaj2002performance}.
YCSB \cite{YCSB} and an extension benchmark----YCSB++ \cite{YCSB++} are proposed to evaluate NoSQL systems for scale-out data services. The metrics are throughput---total operations per second, including reads and writes, and average latency per requests.

PARSEC is  a benchmark suite for studies of Chip-Multiprocessors (CMPs) \cite{BieniaPh.D}. PARSEC includes emerging applications
in recognition, mining and synthesis (RMS) as well as systems applications which mimic large-scale multi-threaded commercial
programs. SPEC CPU2006 provides a snapshot of current scientific
and engineering applications, including  a suite of serial programs that is not
intended for studies of parallel machines \cite{BieniaPh.D}.
SPEC CPU2006 component suite include both CINT2006---the Integer Benchmarks and CFP2006---the Floating Point Benchmarks.
After the benchmarks are run on the system under test (SUT), a ratio for each of them is calculated using the run time on the system under test and a SPEC-determined reference time \cite{SPEC}.

SPEC also proposed a series of benchmarks for Java applications \cite{SPEC}. Among them, SPECjvm2008 is a client JVM benchmark; SPECjbb2005 is a server JVM benchmark, while SPECjEnterprise2010 is a Java enterprise edition application server benchmark. SPECjms2007 is the first industry-standard benchmark for evaluating the performance of enterprise message-oriented middleware servers based on JMS (Java Message Service).  SPECweb2009 emulates users sending browser requests over broadband Internet connections to a web server using both HTTP and HTTPS. It provides banking, e-commerce, and support workloads, along with a new power workload based on the e-commerce workload \cite{SPEC}.
SPECsip\_Infrastructure2011 is designed to evaluate a system's ability to act as a SIP server supporting a particular SIP application \cite{SPEC}.
The application  is modeled after a VoIP deployment 
 \cite{SPEC}.
The metric is the simultaneous number of supported subscribers \cite{SPEC}.

\subsubsection{discussion}
We draw four conclusions in this subsection: first,  there is no systematic work on benchmarking throughput-oriented workloads in data centers. In Section \ref{Characterizing_computing_paradigms}, we have identified three categories of HVC workloads. 
Few previous benchmarks pay attentions to all three categories of applications.

Second, some benchmarking efforts \cite{HiBench} \cite{WL Suite} have focused on MapReduce-based  data analysis applications (belong to our second category of HVC workloads), and the metric is toward evaluating MapReduce  runtime systems. They failed to notice that there are diverse programming models in this field, e.g., Dryad, AllPair \cite{54-Transformer}, since
MapReduce is not a one-size-fits-all solution \cite{54-Transformer}. For example, MapReduce or Dryad is not appropriate for applications
such as iterative jobs, nested parallelism, and irregular parallelism \cite{54-Transformer}.

 Third, a lot of previous benchmarks, e.g., TPC or SPEC efforts,  pay attention to services. Unfortunately, emerging data-intensive services, such as Web search engines are ignored. 

Last, in addition to SPECsip\_Infrastructure2011,  little previous work focuses on interactive real-time applications, which are important emerging workloads, since more and more users tend to use streaming media or VoIP for fun or communications.

\subsection{Our ongoing work on metrics and benchmarks. } \label{our_work}

 Benchmarking is the foundation of evaluating HVC systems.
 However, to be relevant, an HVC benchmark suite needs to satisfy a number of properties as follows:

 First, the applications in the suite should consider a target class of machines \cite{BieniaPh.D}, that is  a data center system designed for throughput-oriented workloads, not a processor or server in this paper.

Second, the HVC benchmark suite should represent three categories of important applications, including services, 
data processing applications, and interactive real-time applications, which is the right target of our ongoing DCBenchmarks project (\url{http://prof.ncic.ac.cn/DCBenchmarks}). In this project, we have released a SEARCH benchmark \cite{Search_Benchmark}. We will release a benchmark for shared data center computer systems running data processing applications soon.

 Third, the workloads in the HVC benchmark suite should be diverse enough to exhibit the range of behavior of the target applications \cite{BieniaPh.D}. Meanwhile, since service providers may deploy different applications, it is important for a service provider to customize their chosen benchmarks relevant to their applications.

 Fourth, no single metric can measure the performance of computer systems on all applications \cite{J.Gray_BenchmarkHandbook}. Since different categories of workloads in HVC have different focuses, we propose different metrics.
For each category of workloads, we propose an aggregate metric which measures a data center system on the whole, and an auxiliary metric to evaluate  energy efficiency, which can be measured on the level of not only  a server but also a data center system.
As shown in Table \ref{Characterizing_computing_paradigms}, for \emph{services}, we propose \emph{requests per minute} as an aggregate metric, and \emph{requests per joule} as an auxiliary metric evaluating energy efficiency. 
 For \emph{data processing applications}, we propose \emph{data processed per minute} as an aggregate metric and \emph{data processed per joule} as an auxiliary metric.
 For interactive real-time applications, we propose {\em maximum number of simultaneous subscribers} as an aggregate metrics and {\em subscribers per watt------a ratio of the maximum number of simultaneous subscribers to the power consumption in a unit time} as an auxiliary metric. A subscribe here can be a user or device.  Due to the space limitation, we will report the experiment results in detail in another paper.

We also developed several innovative performance analysis tools \cite{autoAnalyzer} \cite{DSN_preciseTracer} \cite{TPDS_preciseTracer}, aiding with understanding HVC workloads.

\begin{table*}[hbtp]
\renewcommand{\arraystretch}{1.3}
\caption{Comparison of different benchmarks and metrics.}
\label{Different_Benchmarks_metrics}
\centering
\begin{tabular}{|p{2.6cm}|p{2.8cm}|p{2.2cm}|p{3.5cm}|p{4.2cm}|}
  \hline
  \itshape Benchmark& \itshape Domains &\itshape Level & \itshape Workloads  & \itshape Metrics \\ \hline
   \itshape Linpack \cite{LINPACK} &High performance computing & Super computers & scientific computing code &Float point operations
per second \\ \hline
  \itshape SWap \cite{SUN_SPACE_WATTS} &Enterprise& Systems & Undefined & Performance/(space * watts)   \\ \hline
        \itshape Green 500 \cite{Green500} &High performance computing & Super computers & Scientific computing code & Flops per watt   \\ \hline
  \itshape Graph 500 \cite{betterBechmark} &High performance computing& Super computers & Computing in the field of graph theory &Traversed edges per second (TEPS) \\ \hline
  \itshape JouleSort \cite{Green_Model_Metrics} &Mobile, desktop, enterprise   & Systems & External sort & Records sorted per Joule  \\ \hline
   \itshape SPECpower \_ssj2008 \cite{SPEC} &Enterprise   & Systems & SPECjbb2005 & Ssj\_ops/watt  \\ \hline
                    \itshape  \cite{storageMetrics}             &Storage I/O &Storage systems   &Transaction processing or scientific applications   & I/O or data
rates \\ \hline
  \itshape SPECsfs2008 &Network file systms & File servers & N/a & Operations per second and overall latency of the operations   \\ \hline
    \itshape HiBench \cite{HiBench} &Data-intensive scalable computing & MapReduce runtime systems & Data analysis & Job running time and number of tasks completed per minute  \\ \hline
                \itshape GridMix2 or GridMix3 \cite{GridMix} &Data-intensive scalable computing  & MapReduce runtime systems & Data analysis & Number of completed jobs and running time  \\ \hline
        \itshape WL Suite \cite{WL Suite} &Data-intensive scalable computing & MapReduce runtime systems & Data Analysis &n/a  \\ \hline
                \itshape YCSB \cite{YCSB} or YCSB++ \cite{YCSB++} &Warehouse-scale computing &NoSQL systems & Scale-out data services & Total operations per second and average latency per requests   \\ \hline
                            \itshape PARSEC \cite{BieniaPh.D}& n/a &Chip-Multiprocessors  & Recognition, mining, synthesis, and mimic large-scale multithreaded commercial
programs &n/a   \\ \hline
                \itshape TPC C/E/H \cite{TPC_Benchmarks} &Throughput-oriented workloads & Server systems &  Transaction processing and decision support& Application-specific   \\ \hline
        \itshape SPEC CPU2006 \cite{SPEC} &scientific and engineering applications & Processors & Serial programs & A ratio is calculated using the run time on the system under test and a SPEC-determined reference time   \\ \hline
                \itshape SPECjvm2008, SPECjbb2005, SPECjEnterprise2010, SPECjms2007 \cite{SPEC} &Throughput-oriented workloads &Both hardware and software &Java applications & Throughput (application-specific)   \\ \hline

   \itshape SPECsip \_Infrastructure2011 \cite{SPEC} &Throughput-oriented workloads &Systems &A VoIP deployment & Simultaneous number of supported subscribers   \\ \hline
      \itshape SPECvirt \_sc2010 \cite{SPEC} &Throughput-oriented workloads &Systems &  SPECweb2005, SPECjAppServer2004, and SPECmail2008 & Performance only and performance per watt  \\ \hline
      \itshape SPECweb2009 \cite{SPEC} &Throughput-oriented workloads &Systems & Banking, Ecommerce, and Support & Maximum number of simultaneous user sessions, ratio of the sum of simultaneous user sessions to the sum of watts used  \\ \hline

\end{tabular}
\end{table*}

\section{Conclusion} \label{conclusion}
There are four-fold contributions in this paper. For the first time, we  systematically characterized HVC systems and  identified  three categories of HVC workloads: services, data processing or interactive real-time applications;
we compared HVC with other  computing paradigms  in terms of six dimensions: levels, workloads, metrics, coupling degree, data scales, and number of jobs or service instances;
we widely investigated previous benchmarks, and found there is no systematic work on benchmarking HVC systems;  we presented our preliminary work on HVC metrics and benchmarks.


\section*{Acknowledgment}
We are very grateful to anonymous reviewers. This work is supported by the Chinese 973 project (Grant No.2011CB302500) and the NSFC project (Grant No.60933003).




%

\end{document}